\documentclass[preprint2]{aastex}
\usepackage{spr-astr-addons}
\usepackage{graphicx}
\begin{document}
\title{Large Kinetic Power in FRII Radio Jets}


\shorttitle{Large Kinetic Power in FRII  Radio Jets}
\shortauthors{Ito et al.}

\author{Hirotaka Ito\altaffilmark {1},  Motoki Kino\altaffilmark{2},
    Nozomu Kawakatu\altaffilmark{3},  Naoki  Isobe\altaffilmark{4}, and
    Shoichi Yamada\altaffilmark{1,5} }



\altaffiltext{1}{Science and Engineering, Waseda University, 3-4-1 Okubo,
Shinjuku, Tokyo 169-8555, Japan}
\altaffiltext{2}{ISAS/JAXA, 3-1-1 Yoshinodai, 229-8510 Sagamihara,   Japan} 
\altaffiltext{3}{National Astronomical Observatory of Japan, 181-8588 Mitaka, Japan}
\altaffiltext{4}{Cosmic Radiation Laboratory, 
  Institute of Physical and Chemical Research,
    Wako, Saitama, Japan 351-0198}
\altaffiltext{5}{Advanced Research Institute for Science \&
Engineering, Waseda University, 3-4-1 Okubo,
Shinjuku, Tokyo 169-8555, Japan}




\begin{abstract}
We 
investigate the total kinetic powers ($L_{\rm j}$) and ages 
($t_{\rm age}$) of powerful jets of
  four FR II radio sources (Cygnus A, 3C 223, 3C 284, and 3C 219)
by the detail comparison of the dynamical model
of expanding cocoons with observed ones.
It is found that these sources 
have quite large kinetic powers
with the ratio of $L_{\rm j}$ to the Eddington luminosity ($L_{\rm Edd}$)
 resides in  $0.02 <L_{\rm j}/L_{\rm Edd} <10$.
Reflecting the large kinetic powers,
we also find that the total
energy stored in the cocoon ($E_{\rm c}$)  exceed the 
energy derived 
from the minimum energy condition ($E_{\rm min}$): 
$2< E_{\rm c}/E_{\rm min} <160$. 
This implies that a large amount
 of kinetic power is carried by invisible components such as thermal
 leptons (electron and positron) and/or protons. 
\end{abstract}




\section{Introduction}
\label{intro}

The total kinetic powers of AGN jets 
is one of the most basic physical quantities of the
jet. 
It is however difficult to estimate $L_{\rm j}$,
since most of the observed emissions from AGN jets
are of non-thermal electron origin,
and  the electromagnetic signals
from the thermal and/or proton components is hard to detect.
Hence, the free parameter describing the amount of 
the invisible plasma components always lurks in  
the estimates of $L_{\rm j}$ 
based on the non-thermal emissions.

Recently,  \citet{KK05} (hereafter KK05) proposed
a new  estimate of
 $L_{\rm j}$ for  FRII radio galaxies
 based on the 
dynamical model of cocoon. 
Jets in powerful radio galaxies are 
expected inflate a  cocoon
 into the  surrounding intra-cluster medium (ICM)
 which is overpressured against the ICM \citep{BC89}.
From their model, $L_{\rm j}$ and  $t_{\rm age}$ can be derived
by comparing the cocoon model with the actual 
morphologies of the cocoon based on the radio observations.
However, at the moment, this model  has been
only applied to Cygnus A.
The extension of the number of samples 
are evidently crucial for exploring general characteristics of 
powerful AGN jets. 
For this purpose, we apply the method of KK05
 (with a slight modification)
 to four  FR II radio galaxies  (Cygnus A, 3C 223, 3C 284, and 3C 219).

\section{Cocoon model}
\label{sec:1}

\begin{figure}[h]
\begin{center}
\includegraphics[width=7cm]{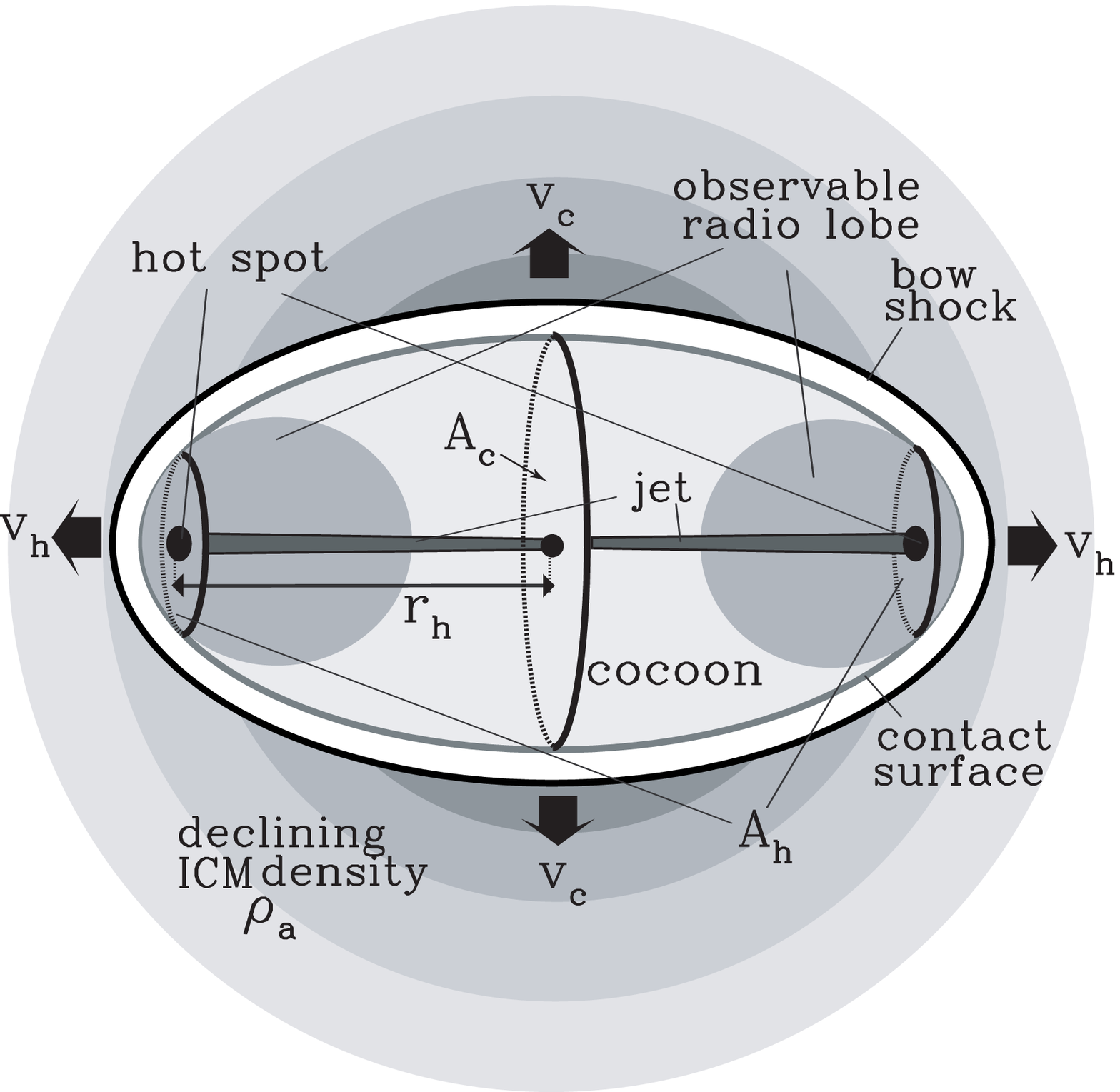}
\end{center}
\caption{{A cartoon of interaction of 
the ICM with declining atmosphere  
and the relativistic jet.}}
\label{cocoon} 
\end{figure}

 Following  KK05,
  we briefly summarize the dynamics of  cocoon 
 in over-pressured regime,
 namely $P_{\rm c} > P_{\rm a}$, where $P_{\rm c}$
and $P_{\rm a}$ are the cocoon pressure 
and the pressure of the ambient ICM, respectively. 
The basic equations which describe 
the equation of motion along jet axis, 
equation of motion of the sideways expansion, 
and the energy equation
are  expressed, respectively, as 
\begin{equation}
\frac{L_{\rm j}}{v_{\rm j}}=
\rho_{\rm a}(r_{\rm h})v_{\rm h}^{2}(t)A_{\rm h}(t),
\label{jet-axis}
\end{equation}
\begin{equation}
P_{\rm c}(t)=
\rho_{\rm a}(r_{\rm c}) \
v_{\rm c}(t)^{2}  ,
\label{lateral}
\end{equation}
\begin{equation}
\frac{d}{dt}
\left( \frac{P_{\rm c}(t)V_{\rm c}(t)}{\hat{\gamma_{\rm c}}-1} \right)
  + P_c(t)\frac{dV_{\rm c}(t)}{dt}
= 2 L_{\rm j}    ,
\label{energy}
\end{equation}
where 
$v_{\rm j}$,
$\rho_{\rm a}$, 
$v_{\rm h}$,
$v_{\rm c}$, $A_{\rm h}$,  and $\hat{\gamma}_{\rm c}$ 
 are 
the velocity of jet,
the density of ambient medium,
 the advance velocity of cocoon head,
 the velocity of  sideways expansion,
 the cross sectional area of cocoon head,
 and the specific heat ratio of plasma inside cocoon, respectively.
 In these equations we also assume that 
 the jet is relativistic, and
 that $L_{\rm j}$ is constant in time.
 $r_{\rm h}(t)=\int_{t_{\rm min}}^{t} v_{\rm h}(t')dt'$, 
 $r_{\rm c}(t)=\int_{t_{\rm min}}^{t} v_{\rm c}(t')dt'$, and
 $V_{\rm c}(t)= \frac{4}{3}\pi r_{\rm c}(t)^2 r_{\rm h}(t)$
 are 
the distance from the jet apex to the hotspot,
 the radius of the cocoon body, and 
 the volume of the cocoon
 where $t_{\rm min}$ is the initial time of source evolution.
 The declining  mass density 
 of ICM $\rho_{\rm a}$  is given by
 $\rho_{\rm a}(r) ={\bar\rho}_{\rm a}(r/{\bar r}_{\rm h})^{-\alpha}$
 where ${\bar r}_{\rm h} \equiv r_{\rm h}(t_{\rm age})$. 
 A cartoon of the cocoon model is illustrated in Fig. \ref{cocoon}.
 We slightly improve the model of KK05
 as follows;
 (i) the geometrical factor of $V_{\rm c}$ is more accurate, and 
 (ii) the inclusion of $PdV$ work done by cocoon.

 We  further assume
 the lateral expansion of the
 cocoon as $A_{\rm c}(t) = {\bar A}_{\rm c}  (t/t_{\rm age})^{X}$,
 where $A_{\rm c}(t) = \pi r_{\rm c}(t)^2$ is
 the cross sectional area of cocoon body. 
 By assuming that all physical quantities also have 
 a time-dependence of form  $A={\bar A} ~ (t/t_{\rm age})^{Y}$,
  $v_{\rm c}$ is determined as
\begin{eqnarray}\label{v_c}
v_{\rm c}(t)=
{\bar v}_{\rm c}
\left(\frac{t}{t_{\rm age}}\right)^{0.5X-1}
=
\frac{{\bar A}_{\rm c}^{1/2}}{t_{\rm age}}
{\cal C}_{\rm vc}
\left(\frac{t}{t_{\rm age}}\right)^{0.5X-1} .
\end{eqnarray}
 From this, we obtain the following expressions:
%
\begin{eqnarray}\label{Pbar}
{\bar P}_{\rm c} \equiv
P_{\rm c}(t_{\rm age})=
{\bar\rho}_{\rm a} {\bar v}_{\rm c}^{2}
{\cal C}_{\rm pc}
\left(\frac{{\bar v}_{\rm c}}{v_{0}}\right)^{-\alpha},
\end{eqnarray}
\begin{eqnarray}\label{v_hbar}
{\bar v}_{\rm h}\equiv
v_{\rm h}(t_{\rm age})=
\frac
{L_{j}}
{{\bar\rho}_{\rm a}{\bar v}_{c}^{2}{\bar A}_{\rm c}}
{\cal C}_{\rm vh} 
\left(\frac{{\bar v}_{\rm c}}{v_{0}}\right)^{\alpha},
\end{eqnarray}
\begin{eqnarray}\label{Ahbar}
{\bar A}_{\rm h} \equiv
A_{\rm h}(t_{\rm age})=
\frac{L_{\rm j}}
{v_{\rm j}{\bar\rho}_{\rm a} 
{\bar v}_{\rm h}^{2}}
{\cal C}_{\rm ah}
\left(\frac{{\bar v}_{\rm h}}{v_{0}}\right)^{\alpha},
\end{eqnarray}
 where 
 ${\cal C}_{\rm vh}= \frac{3}{2}
 (\hat{\gamma}_{c}-1)(0.5X)^{-\alpha}[3-(2-0.5\alpha)X]/[4-(1-0.5\alpha)X]$,
 ${\cal C}_{\rm vc}=0.5X/ \pi^{1/2}$,
 ${\cal C}_{\rm pc}=(0.5X)^{\alpha}$, 
 ${\cal C}_{\rm ah}=[X(-2+0.5\alpha)+3]^{-\alpha}$, and
 $v_{0}\equiv {\bar r}_{\rm h}/t_{\rm age}$. 
 Note that the only  difference from KK05 is
 ${\cal C}_{\rm vh}$.
%
 Although it is not dealt in the present study,
 the main difference from the previous models
 \citep{BDO97, KA97} is that
 these solutions can describe  non-self-similar evolution.
 Note that when self-similar condition is imposed,
 time dependence in the physical quantities becomes identical
 to these previous models.

Throughout this paper, we assume  $\hat{\gamma}_{\rm c} = 4/3$
since  cocoon is expected to be dominated by relativistic particles
\citep{KKI07}.
We also assume that the aspect ratio of the cocoon 
 (${\cal R}\equiv r_{\rm c} /r_{\rm h}$)
to be constant in time (self-similar evolution).
 Since ${\cal R}(t) \propto t^{X(2.5-0.5\alpha)-3}$,
 this assumption leads to $X = 6/(5 - \alpha)$.

\section{Comparison with the observation and the model}
\label{sec:2}
 By substituting the observed
 values of ${\bar r}_{\rm h}$, ${\bar r}_{\rm c}$ (or
 ${\cal R}$),  ${\bar A}_{\rm h}$, 
 $\bar{\rho}_{\rm a}$, and $\alpha$  in Eq. (\ref{Pbar}) -
 (\ref{Ahbar}), $L_{\rm j}$, $t_{\rm age}$, and also ${\bar P}_{\rm c}$
   are determined uniquely.
 In this section, we explain how to extract these quantities from the 
 observations 
 and show the obtained $L_{\rm j}$ and $t_{\rm age}$

\begin{figure*}
\includegraphics[width=17cm]{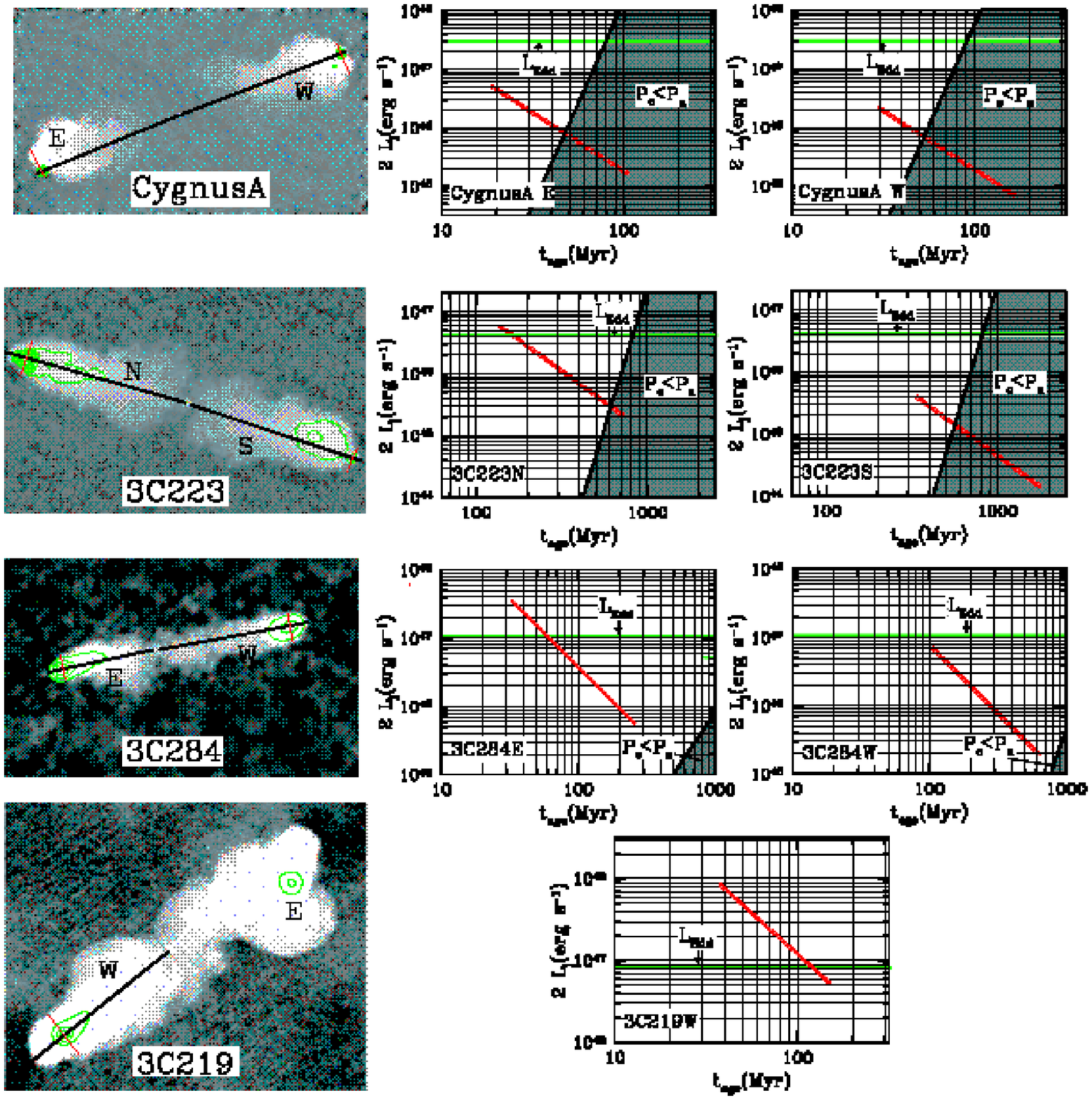}
\caption{{\it left panels}: Logarithmic-scaled 5-GHz VLA map of Cygnus A 
  \citep{PDC84} and 1.5GHz VLA maps of
  3C 223 \citep{LP91}, 3C 284 \citep{LPR86}, and  
 3C 219  \citep{CBBP92}
 with linearly spaced contours (green lines) are displayed.  
 The straight 
 lines overlaid in each map denote the lines we have used
 to measure $r_{\rm h}$ (black lines) and $A_{\rm h}$ (red lines).
 {\it right panels}: 
 The red lines show the obtained range of $L_{\rm j}$ and $t_{\rm age}$.
 The shaded regions show forbidden ranges where the overpressure
 condition ($P_{\rm c} > P_{\rm a}$) is not satisfied.
 Also  the Eddington luminosities (green lines) are displayed for
 comparison.}
\label{radiopower}      
\end{figure*}

In the left panels of  Fig. \ref{radiopower}, we show the
VLA radio images of 
Cygnus A,
3C 223, 
3C 284, and 
3C 219  in logarithmic scale. 
 Contours in linear scale (green lines) are also displayed for the purpose to
 determine the position of hotspot accurately.
 The overlaid straight lines which crosses perpendicular to each 
 others at the hotspot are the lines we used to measure 
$r_{\rm h}$ (black lines) and $A_{\rm h}$ (red lines). 
 Here, $A_{\rm h}$ is measured as the cross sectional
 area of the radio lobe at the position of the hotspot.  
Contrary to $A_{\rm h}$,
 it is difficult to measure $r_{\rm c}$ 
from the VLA radio images
since the  cocoon emission coming from 
the cross section perpendicular
to the AGN core location is very dim at GHz radio frequency 
because of the synchrotron cooling.
Therefore, we treat $r_{\rm c} = {\cal R}r_{\rm h}$ as a free parameter.
In order to take account of large ambiguity,
 we examine sufficiently wide range of $0.5<{\cal R}<1$.
In the present study, we neglect projection effect on $r_{\rm h}$ for
simplicity.
Under the assumption of self-similar evolution, 
$L_{\rm j}$ and $t_{\rm age}$ depend 
on $r_{\rm h}$  as
$L_{\rm j}\propto r_{\rm h}^{-4}$ and
$t_{\rm age}\propto r_{\rm h}^3$, respectively.
However, since the angle to the line of sight is not expected
to  deviate largely from $\pi / 2$ in radio galaxies,
this effect does not cause significant change in our results.

The adopted values of $\rho_{\rm a}$, $\alpha$, and
$P_{\rm a}$ are based on  the literatures of 
\citet{RF96, SW02} for Cygnus A,
\citet{CB04} for 3C 223 and 3C 284,
and \citet{HW99}  for 3C 219, respectively. 
 Note that
 since
 most of the
 radio sources show asymmetries in their shape
  among the pair 
 of the lobes (Fig. \ref{radiopower}),
 we analyze each side of the lobe independently.
 However,
 we only analyze jet in the western side for 3C 219
 since the eastern lobe showed severe deformation.

The resultant $L_{\rm j}$ and $t_{\rm age}$ 
are displayed in the right panels of  Figs. \ref{radiopower}.
 The red  lines displayed in the figure are 
the resultant range of $L_{\rm j}$ and $t_{\rm age}$, 
 and their ranges reflect the uncertainty in $\cal{R}$. 
 The age and power depend on the aspect ratio $\cal{R}$ as 
 $t_{\rm age} \propto \cal{R}$$^{ 4 - \alpha}$  and
 $L_{\rm j} \propto \cal{R}$$^{ 2 \alpha - 8}$,
 and, therefore,  satisfy 
 $L_{\rm j} \propto t_{\rm age}^{ -2 }$. 
 Since $\alpha$  does not exceed $4$ in any of four sources,
 a lower aspect ratio corresponds to
 a higher power with a lower age.
 The range   included in the 
 shaded region is the forbidden region since
 the overpressure condition is violated. 
 The Eddington luminosity ($L_{\rm Edd}$) of each source, 
 is also shown (green line) in these figure for comparison.
 The black hole masses ($M_{\rm BH}$)
 are taken from \citet{TMA03} for Cygnus A,
 \citet{MCF04} for 3C 223 and 3C 219. 
 For 3C284,   
  we derive $M_{\rm BH}$ from the B-band magnitude
 of  host galaxy \citep{SRH05} by using the empirical
 correlation of B-band magnitude and black hole mass \citep{MCF04}. 
 In Table \ref{table3}, we summarize 
 the allowed values of $L_{\rm j}$ and $t_{\rm age}$
 and  the other relevant physical properties of the cocoon.
 Reflecting the asymmetries in their shape,
 the obtained $L_{\rm j}$ and $t_{\rm age}$ show discrepancy among
 the pair of lobes
  especially in 3C 223 and 3C 284, and 3C 219.
 Since it seems natural to suppose that
 the properties of the jets are intrinsically
symmetric and equal power and age on
both side, 
we expect that  the discrepancy is 
due to the asymmetry and/or inhomogeneity of ICM density profile. 
Here we interpret that the actual values of $L_{\rm j}$ and $t_{\rm age}$ is
in the range obtained from both lobes.

%

%

\begin{table*}
\begin{center}
\caption{The obtained properties of the jet and the cocoon together
 with minimum energy of the radio lobe.\label{table3}}
\begin{tabular}{lccccccccccccccllllll}
\hline\noalign{\smallskip}
 Source & $L_{\rm j}$ 
       & $t_{\rm age}$ 
       & $M_{\rm BH}$
       & $2 L_{\rm j}/L_{\rm Edd}$ 
       & $E_{\rm c}$ 
       & $E_{\rm min}$
       & $\eta_{\rm min}$
 \\

  & ($10^{46}$ ergs s$^{-1}$)
       & (Myr)
       & (M$_{\odot}$)
       &
       & ($10^{60}$ ergs) 
       & ($10^{60}$ ergs)
       &    \\

\noalign{\smallskip}\hline\noalign{\smallskip}
 Cygnus A E & 0.4 - 2.6 & 19 - 47   &
      2.5$\times$10$^9$ &
      0.025 - 0.16 &
      3.4 - 8.8  &
       1.4  
      & 2.3 - 6.1   
      \\

 Cygnus A W & 0.35 - 1.1 & 30 - 53  &
      2.5$\times$10$^9$ &
      0.021 - 0.068 &
      3.2 - 5.7  &
      1.4  
      & 2.2 - 4.0 
      \\

 3C 223 N & 0.15 - 2.9 & 140 - 610  &
      3.2$\times$10$^8$ &
      0.072 - 1.43 & 
      16 - 70  &
      0.88 
      & 18 - 79
      \\

 3C 223 S & 0.071 - 0.2 & 330 - 560  &
      3.2$\times$10$^8$ &
      0.034 - 0.097 & 
      6.9 - 12  &
      0.88    
      & 7.8 - 13
      \\

3C 284 E  & 0.3 - 18 & 32 - 260  & 
      8.2$\times$10$^8$ &
      0.053 - 3.4 & 
      14 - 110  &
       1.8    
      & 7.7 - 62      \\

 3C 284 W &  0.1 - 3.6  & 100 - 630  &
      8.2$\times$10$^8$ &
      0.018 - 0.67 & 
      11 - 68  &
      3.0    
      & 3.7 - 23
      \\

3C 219 W  & 2.6 - 43 & 37  - 150  &
      6.3$\times$10$^8$ &
      0.65 - 10 &
      63 - 250  &
      1.6  
      & 40 - 160      \\

\noalign{\smallskip}\hline
\end{tabular}
\end{center}
\end{table*}

\section{Discussions}
\subsection{On $L_{\rm j} / L_{\rm Edd}$}
In Table \ref{table3},
 the total kinetic power of the jet
 normalized by the corresponding Eddington luminosity, 
 $2 L_{\rm j} / L_{\rm Edd}$, is
displayed. 
It can be seen that $2 L_{\rm j} / L_{\rm Edd}$ takes 
quite high value ranging from $\sim 0.02$ to $\sim 10$. 
Postulating that the relativistic jet emanating 
from the AGN is powered by some part of released 
gravitational energy 
  of the accreting matter (e.g., Marscher et al. 2002),
 these values give the minimum mass accretion rate
normalized by Eddington mass accretion rate.
Interestingly, our results indicate that 
mass accretion rates are super-Eddington ones
in some FR IIs (3C 219 and 3C 284) since 
$2 L_{\rm j} /L_{\rm Edd} \simeq 1$. 

\subsection{On the plasma content}
It is intriguing to explore 
how much amount of the internal energy,
 $E_{\rm c} = P_{\rm c}V_{\rm c}/(\hat{\gamma}_{\rm c} - 1)$, 
is deposited in the cocoon 
compared with the widely-discussed minimum energy, 
$E_{\rm min}$, of the radio lobe obtained from the minimum energy
condition for the non-thermal electrons and magnetic fields.
Here we calculate $E_{\rm min}$ based on the 
 observation of 178MHz band radio emission \citep{HAP98}.
 In Table \ref{table3} we summarize
 the resultant $E_{\rm min}$, 
  $E_{\rm c}$, and $\eta_{\rm min}$  which we define as
 the fraction of  $E_{\rm c}$ to  $E_{\rm min}$
  ($\eta_{\rm min} \equiv E_{\rm c}/E_{\rm min} $).
 In all sources $E_{\rm c}$  exceeds $E_{\rm min}$, and 
 the range of the ratio is obtained as  $ 2< \eta_{\rm min} <160$.
 This implies that minimum energy condition is unlikely to
 be realized in these sources.

 Using $\eta_{\rm min} $,
 here we  investigate the 
 plasma content in the cocoon.
 Considering the components of energy, $E_{\rm c}$ is
 sum of the energy of
 the non-thermal  
 leptons (electron/positron) and, if present, the energy of the 
 unobservable particles such
 as thermal leptons and/or protons.
 The  large excess energy of $E_{\rm c}$ from $E_{\rm min}$
 is due to contributions from either of these components.

 Let us consider  the possibility of large contributions  from
 the non-thermal leptons.
 In this case $\eta_{\rm min}$ is given as
 $\eta_{\rm min} = U_{e}^{\rm NT}/U_{\rm min}$
 where 
 $U_{e}^{\rm NT}$ and
  $U_{\rm min}$ are
 the energy  density the nonthermal leptons  and
 the minimum energy density
 ($U_{\rm min} \equiv E_{\rm min}/V_{\rm c}$), respectively.
 It is useful to express $U_{e}^{\rm NT}/U_{\rm min}$
 in terms of $U_{e}^{\rm NT}/U_{\rm B}$, where
  $U_{\rm B}$ is the energy
  density  the magnetic field
 since
 $U_{e}^{\rm NT}/U_{\rm B}$ is widely investigated by a lot of authors
 (e.g.,  Kataoka and Stawarz; Croston et al. 2005).
 In the case of power-law distributed leptons, 
 synchrotron luminosity ($L_{\nu}$)  is determined as
 $L_{\nu}  \propto U_{\rm e}^{NT}U_{\rm B}^{4/3}V_{\rm c}$.
 Hence with $L_{\rm \nu}$ and $V_{\rm c}$ observed, $U_{\rm e}$
 and $U_{\rm B}$ are not independent of each other, and
 $U_{\rm e}^{\rm NT}/U_{\rm min}$ can be written
 as
 $U_{\rm e}^{\rm NT}/U_{\rm min} =0.5 (U^{\rm NT}_{\rm e}/U_{\rm
 B})^{3/7}$.
 Since recent studies
 shows that
 $1<U_{\rm e}^{\rm NT} / U_{\rm B} <10$  on average,  
the value of $\eta_{\rm min}$ is expected to be up to $\sim 1$ at most.
Thus,  
non-thermal  leptons are unlikely to be  the main carrier of the energy.
We conclude that  significant amount of energy 
 is carried by invisible components such as thermal
 leptons (electron and positron) and/or protons.

\section{Summary}

In this paper we have investigated the 
total kinetic power and the age of the relativistic jet
in four FRII radio galaxies
(Cygnus A, 3C 223, 3C 284, and 3C 219).
Below we summarize our main results.

(I)
 A large fraction of Eddington power 
in the range of $0.02 - 10$ is carried away 
as a kinetic power of the jets in the FR II sources.

(II)
 The energy deposited in the cocoon, $E_{\rm c}$,
 exceeds the minimum energy, $E_{\rm min}$,
 by a factor of 
 $2 - 160$.

\begin{acknowledgements}

 This work was partially supported by the Grants-in-Aid for the
 Scientific Research (14740166, 14079202) from Ministry of Education,
 Science and Culture of Japan and by Grants-in-Aid for the 21th century
 COE program ``Holistic Research and Education Center for Physics of
 Self-organizing Systems''.
 This research has made use of SAOimage DS9, developed by Smithsonian
 Astrophysical Observatory.
\end{acknowledgements}




\end{document}